\documentclass[floatfix]{revtex4}

\usepackage{graphicx}
\begin{document}
% You should use BibTeX and revtex.bst for references
\bibliographystyle{revtex}

% number  and your conference paper identification number on the
% title page in preprint mode. Multiple \preprint commands are allowed.
%\preprint{}

\title{ADD extra dimensional gravity and di-jet production at hadron 
colliders}
% Optional argument for running titles on pages
%\title[]{}



\author{M.A. Doncheski}
\email[]{mad10@psu.edu}
%\thanks{}
\affiliation{Department of Physics, Pennsylvania State University, 
Mont Alto, PA 17237 USA}


\date{\today}

\begin{abstract}
We re-analyze dijet production at hadron colliders (the Tevatron at Fermilab 
and the Large Hadron Collider, LHC, at CERN), to determine the potential 
limits on Planck mass in ADD type extra dimensional gravity scenarios.  We try 
a variety of experimental observables in order to maximize the exclusion 
limits; we find that the $p_{_T}$, $p_{_T}^2$ and $\tau$ distributions give 
the highest search limits, and these observables provide comparable reaches.
\end{abstract}
% \pacs{}

%\maketitle must follow title, authors, abstract and \pacs
\maketitle

%\section{Introduction}

\section{Introduction}

Conventional wisdom tells us that gravity is the weakest, by far, of the four 
fundamental forces of the universe.  However, the possibility of large extra 
dimensions \cite{add,rs} where gravity becomes strong at scales of order a 
TeV, may lead to a complete revision of conventional wisdom.  This possibility 
has spawned a great deal of research, both phenomenological and experimental, 
into the discovery or exclusion of extra dimensional gravity scenarios.

The first such scenario \cite{add}, commonly known as ADD extra dimensional 
gravity, suggests that extra spatial dimensions (the bulk) beyond the usual 3 
(the wall) exist in which gravity operates.  For distances larger than the 
extra dimension length scale, the effective Planck mass is large, 
$M_P \sim 10^{19} \; GeV$, while for distances smaller than the extra 
dimension length scale, the true Planck mass is small, 
$M_S \sim {\cal O}(1 \; TeV)$.  The attractiveness of ADD and other extra 
dimensional gravity scenarios is that they solve the hierarchy and naturalness 
problems by moving the scale of gravity to something near the electroweak 
scale.

The phenomenology of the ADD model has been studied extensively; the Feynman 
rules are given in Ref.~\cite{pheno}.  Among the many processes studied to 
date, Atwood, Bar-Shalom and Soni \cite{dijet} recently studied the effect 
of graviton tower exchange to dijet production at hadron colliders.  Unlike a 
direct graviton production process, where the graviton produced appears as 
a missing $E_T$ signature, dijet production is sensitive to virtual graviton 
exchange in the $s$, $t$ and/or $u$ channels (depending on the subprocess).  
Virtual graviton exchange can modify various experimental observables to be 
significantly different than Standard Model (SM) predictions, and ADD 
scenarios can be discovered or excluded based on measured deviations from SM 
predictions.  Similar analyses are possible under other extra dimensional 
gravity scenarios, such as Randall-Sundrum \cite{rs}, but those analyses are 
beyond the scope of this study.

\section{Calculation}

In Ref.~\cite{dijet}, equations for all the necessary parton level 
subprocesses are given.  The authors of Ref.~\cite{dijet} reported exclusion 
limits at the Tevatron and LHC based on deviations from the SM $\tau$ 
distribution, where $\tau$ is the usual product of parton momentum 
fractions
\begin{equation}
\tau = x y = \frac{M_{jj}^2}{s}
\end{equation}
but can also be expressed in terms of experimental observables $M_{jj}$ (the 
jet-jet invariant mass) and $s$ (the square of the center of mass energy).

Based on recent compositeness searches by the CDF and D0 \cite{expt} 
collaborations, we felt that alternate experimental observables could improve 
the search reach here.  One favorite observable is transverse momentum; 
$p_{_T}$ and $p_{_T}^2$ are natural choices.  Psuedorapidity,
\begin{equation}
\eta = \log{\frac{1 + \cos \theta}{1 - \cos \theta}}
\end{equation}
is another commonly used observable.  Related to $\eta$ is
\begin{equation}
\chi = \frac{1 + \cos \theta}{1 - \cos \theta}.
\end{equation}
In addition, a ratio of $M_{jj}$ distribution with $\eta > \eta_0$ to $M_{jj}$ 
distribution with $\eta < \eta_0$ was found to be useful in compositeness 
searches.

In order to simulate detector acceptance, we count jets only when 
$| \eta | < 1$ and $p_{_T} > 10 \; GeV$.  Furthermore, we assume an integrated 
luminosity of $2 \; fb^{-1}$ for the Tevatron and $30 \; fb^{-1}$ for the 
LHC.  With these acceptance cuts and integrated luminosities, the event 
rates are large, leading to a high level of sensitivity to deviations from the 
SM predictions.  As will be clear below, this analysis will not depend 
strongly on the value of the $p_{_T}$ cut; the strongest deviation from SM 
occurs at $p_{_T}$ significantly higher than $10 \; GeV$.  For our analysis, 
CTEQ5M \cite{cteq} distributions are used.

For all observables, a $\chi^2$ analysis was performed, where
\begin{equation}
\chi^2 = \sum \left( \frac{{\cal N}_i - {\cal N}_i^{\rm{SM}}}
                   {\delta {\cal N}_i^{\rm{SM}}} \right)^2
\end{equation}
where ${\cal N}_i$ is the event number in a specific bin, and only statistical 
errors were considered, so that the uncertainty in ${\cal N}_i$, 
$\delta {\cal N}_i$, equals $\sqrt{{\cal N}_i}$.  $\chi^2 = 4$ corresponds to 
a 95\% C.L. deviation from the SM.  We chose the number of bins to be 50 for 
the Tevatron and 100 for the LHC; this corresponds to $p_{_T}$ bin sizes of 
$20 \; GeV$ and $70 \; GeV$, respectively.

\begin{figure}
\includegraphics[width=3.0in,angle=-90]{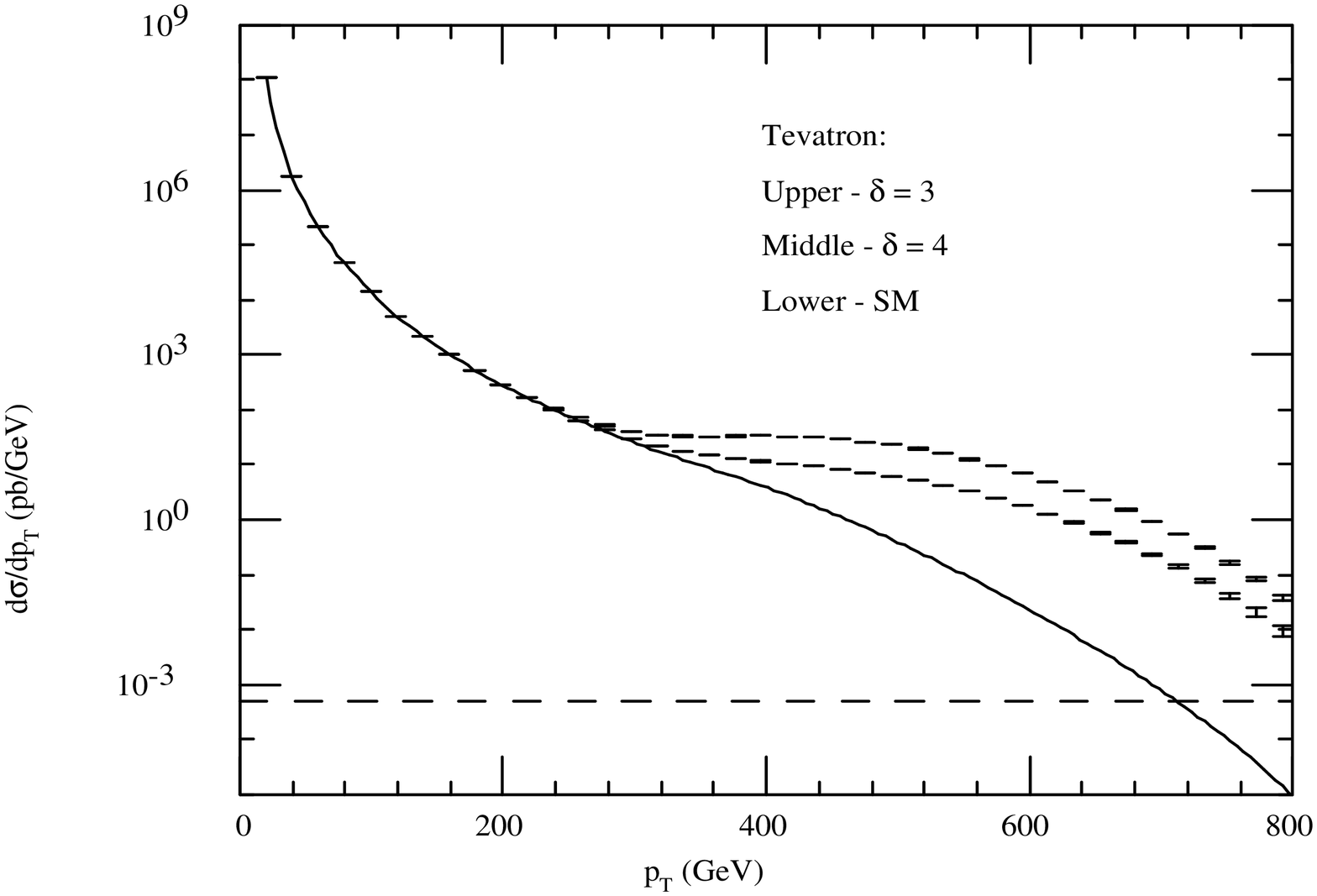} \\
\includegraphics[width=3.0in,angle=-90]{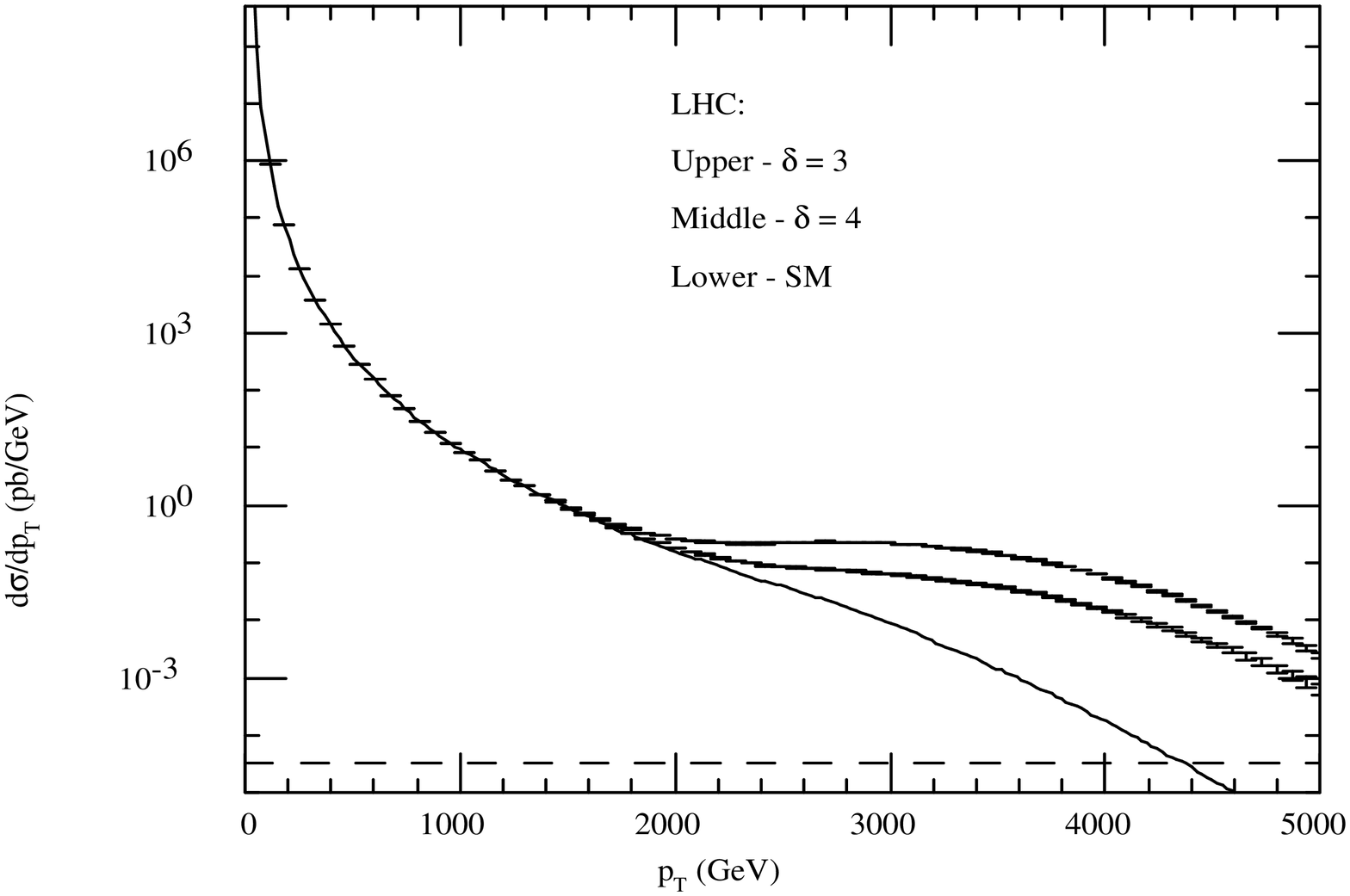}%
\caption{(a) $d \sigma /d p_{_T}$ vs. $p_{_T}$ for the for the SM (solid 
lines) and with extra dimensional gravity ($\delta = 3, 4$ as indicated on the 
figures) for the Tevatron and LHC respectively.  The error bars correspond to 
$1 \; \sigma$ statistical uncertainties only; an integrated luminosity of 
$2 \; fb^{-1}$ ($30 \; fb^{-1}$) and $M_S = 1.0 \; TeV$ ($7.0 \; TeV$) is 
assumed for the Tevatron (LHC).  The horizontal dashed line indicates 1 event 
per bin.}
\label{Fig1}
\end{figure}

A comparison of the $p_{_T}$ distribution, $d\sigma/dp_{_T}$, is shown in 
Figure~\ref{Fig1}, for both the Tevatron and the LHC; the SM (solid line) is 
compared with extra dimensional gravity predictions for the number of extra 
dimensions $\delta$ of 3 and 4, as indicated on the figures.  The extra 
dimensional gravity points include statistical uncertainty only, and the 
horizontal dashed line indicates 1 event/bin.  The other extra dimensional 
gravity parameter, the Planck mass $M_S$, is chosen to be $ 1 \; TeV$ for the 
Tevatron and $7 \; TeV$ for the LHC.  There is a large excess of events at 
high $p_{_T}$, and it is clear that the exclusion limits possible far exceed 
the values of $M_S$ chosen to produce Figure~\ref{Fig1}.

The search/exclusion limits possible at the Tevatron are $M_S = 3.1 \; TeV$ 
($2.6 \; TeV$) for $\delta = 3$ (4), while the limits at the LHC are 
$20.8 \; TeV$ ($17.4 \; TeV$).  Similar limits are possible using $p_{_T}^2$ 
and $\tau$ distribution; the limits for these observables are only a few 10s 
of $GeV$ lower.  The other observables mentioned above ($\chi$, $\eta$, 
$M_{jj}$, {\it etc.}) provide limits that are significantly lower than those 
from the $p_{_T}$ distribution.  The authors of Ref.~\cite{dijet} chose one 
of the best possible observables on which to base their analysis.

The exclusion limits reported here are slightly higher than those in 
Ref.~\cite{dijet}, even for the $\tau$ distribution.  This is almost certainly 
due to our use of a finer binning of observables in the distributions.  Just 
as a distribution will give a higher $\chi^2$ than a total cross section, a 
more finely binned distribution will give a higher $\chi^2$ than a coarser 
binning, assuming sufficient event numbers.  The binning used here for the 
$p_{_T}$ distribution is rather coarse, so our results are, in a sense, 
conservative.

\section{Conclusions}

Due to the extremely high event rate, dijet production at hadron colliders is 
a favorite for searches for Physics Beyond the Standard Model.  As shown 
originally by the authors of Ref.~\cite{dijet}, dijet production at the 
Tevatron and the LHC is very sensitive to virtual graviton exchange effects, 
as provided by extra dimensional gravity scenarios.

In this re-analysis, we studied a number of observables related to dijet 
production, and found that the $p_{_T}$ distribution was more sensitive to 
virtual graviton effects than the $\tau$ distribution proposed by the authors 
of Ref.~\cite{dijet}, but only slightly more sensitive.  A more thorough 
analysis, including a more realistic detector simulation and systematic 
effects, is required, but the more careful analysis will not change the fact 
that dijet production at hadron colliders will be an important probe of 
extra dimensional gravity models.

\begin{acknowledgments}
The author thank Tom Rizzo and Greg Landsburg for many helpful conversations 
and communications.  This research was supported in part by the Commonwealth 
College and the Eberly College of Science of Penn State University.
\end{acknowledgments}

%\bibliography{ncqed.refs}

\end{document}